REVIEW

# Success of alignment-free oligonucleotide (k-mer) analysis confirms relative importance of genomes not genes in speciation and phylogeny

Donald R. Forsdyke

Corresponding author: Donald R. Forsdyke, Department of Biomedical and Molecular Sciences, Queen's University, Kingston, Ontario, Canada K7L3N6. Tel.:01-613-533-2980; E-mail: forsdyke@queensu.ca

Running header: *Alignment-free phylogenomics*

**Donald R. Forsdyke** is a professor in biochemistry at Queen's University and author of *Evolutionary Bioinformatics*. Allied interests are biohistory, evolutionary biology, genetics and immunology. He has written scientific biographies of George Romanes (Darwin's research associate), and William Bateson (geneticist).

*Main Headings*

    **Abstract**
    **Introduction**
    **Microorganisms as species**
    **Ad hoc DNA-based classification methods**
    **Theoretical link with speciation**
    **Molecular link with speciation**
    **Nucleic acids versus proteins**
    **Concluding remarks**

**Words: 8101**
**Figures and Tables: 0**
**References: 101**

**Abstract**

The utility of DNA sequence substrings (k-mers) in alignment-free phylogenetic classification, including that of bacteria and viruses, is increasingly recognized. However, its biological basis eludes many twenty-first century practitioners. A path from the nineteenth century recognition of the informational basis of heredity to the modern era can be discerned. Crick's DNA "unpairing postulate" predicted that recombinational pairing of homologous DNAs during meiosis would be mediated by short k-mers in the loops of stem-loop structures extruded from classical duplex helices. The complementary "kissing" duplex loops – like tRNA anticodon-codon k-mer duplexes – would seed a more extensive pairing that would then extend until limited by lack of homology or other factors. Indeed, this became the principle behind alignment-based methods that assessed similarity by degree of DNA-DNA reassociation in vitro. These are now seen as less sensitive than alignment-free methods that are closely consistent, both theoretically and mechanistically, with chromosomal anti-recombination models for the initiation of divergence into new species. The analytical power of k-mer differences supports the theses that evolutionary advance sometimes serves the needs of nucleic acids (genomes) rather than proteins (genes), and that such differences have often played a role in early speciation events.

**Introduction**

The utility of alignment-free sequence analyses in the classification of, and determination of evolutionary relationships between, living forms, is endorsed by the National Center for Biological Information (NCBI). In February 2019 it announced its adoption of average nucleotide identity (ANI) analysis to correct existing taxonomic information in its databases [1]. This marks a major step in the shift from distinctions based entirely on phenotypes to those based on the bioinformatic analysis of genomes. With increased input from information scientists, a new science – evolutionary bioinformatics – is emerging [2].

However, when it comes to "analytical efficiency" a "gap between theory and practice" is noted. Given the exponential growth of sequence data-banks, there is call for more "computational pragmatism." Novel alignment-free approaches are disparaged on the grounds that they "lack any biological intuition" and "altogether discard the concept of homology," which plays a key role in pairwise genome comparisons [3]. Thus, three major questions are identified. Are alignment-free approaches valid? Are they applicable to all biological forms, including those where species classifications are in doubt? Are they relevant to anything fundamental in biology?

The practical advantages of alignment-free methods over alignment-based methods are now clear [4,5]. This is particularly evident with retroviruses whose high recombination rates result in frequent sequence disruptions [6]. Indeed, the NCBI endorsement reflects a growing consensus on the validity of comparing oligonucleotide (k-mer) frequencies *irrespective of their relative order*. However, the validity question cannot be detached from the other questions. Establishing a biological relationship further validates. It is claimed that, when k>1, "individual k-mers can be viewed as embodying parts of the homology signal in a sequence," and that "alignment-free methods have not abandoned models or homology and can be biologically intuitive" [7]. But this case has yet to be fully made. Indeed, Zieleninski *et al*. [4] lament that: "The absence of well-defined benchmarks covering various evolutionary scenarios of sequence divergence creates a major obstacle for researchers who simply need to know the current 'best' tool."

Having examined nucleotide sequences pragmatically, in 1986 an early exponent of the new approach did not speculate on a possible biological basis for his observations apart from "intuitive biological reasoning" [8]. Many came to assume that "most determinations reveal nothing fundamental about the biology of the organisms involved" [9]. Even Erwin Chargaff's observation in 1951 [10] that genomes have species-specific GC% values (his 'GC% rule'), which became *de rigueur* to include in species descriptions, was dismissed: "Base distribution (GC content) is recognized to be an attribute of many organisms' classification, but to have no fundamental meaning other than an expression of the base complementarity rules" [9].

Thus, from the outset twenty-first century practitioners of alignment-free approaches were uncertain as to an underlying theoretical basis for their work. In 2004 Teeling *et al*. [11]

pondered "the evolutionary significance of species-specific patterns that are observed." A 1995 report that different human chromosomes have the same k-mer patterns [12], was extended in 2005 by Dehnert *et al*. [13], who wondered "which mechanism synchronizes the correlation pattern of chromosomes leading to this remarkable degree of similarity within the chromosomes of a species." In 2006 van Passell *et al*. [14] regretted "our lack of understanding of the factors that shape the nucleotide composition." In 2009 Richter and Rosselló-Móra [15] noted that "oligonucleotide frequencies carry a species-specific signal, but the evolutionary reasons behind this have not been comprehensive explained so far."

As I have reviewed elsewhere [2, 16-18], species arise from the divergence of the nucleic acid sequences of preexisting species. Traces of early divergence-initiating and sustaining events may have been retained in the species we study today, and may have been revealed by alignment-free approaches. The case for these 'echoes from the past' is made here in four steps. Since much of the early work was based on short sequences from small microbial genomes, the growing acceptance that microbes are classifiable in the same way as eukaryotes, is first considered. Next, various genome-based methods of species classification and phylogeny are contrasted. The new alignment-free methodology is then related to various theoretical and molecular aspects of speciation. Finally, attention is drawn to the emerging appreciation that evolutionary advance serves the needs of nucleic acids (genomes) rather than of genes and their products (e.g. proteins). This contrasts with a narrowly conceived view of natural selection as calling for responses of organisms to aspects of their environments that do not include the nucleic acids of members of their own or allied species. Just as "ecotypes" occupy ecological niches [19], so species can be regarded as occupying recombinational niches. There is a conventional phenotype and a genome phenotype.

## Microorganisms as species

The historical definition of a species as a group of organisms that sexually reproduce among themselves but not with members of other species [20], has long appeared problematic for organisms that prominently reproduced asexually. These include bacteria, archaea and viruses. Thus, Hunter [21] argues: "Not only are viruses technically not cells at all, lacking all the protein machinery, but they can also not be categorized as species on the basis of reproductive isolation since they depend on their host for replication." Furthermore, while conceding that asexual organisms "form isolated and separate populations that are phenotypically and genotypically distinct from those of other species," a taxonomic authority declares: "Reproductive isolation is commonly used to delineate species boundaries of animals, plants and fungi in the classical biological species concept, but this property cannot be used for virus species delineations because they do not interbreed in any conventional sense" [22]. Thus, viruses have been formally delineated by "multiple criteria" that include morphology, replication properties, and host range [23].

The increased reliance on genotypic, rather than phenotypic, criteria for taxonomic identification to be detailed below, has progressively eroded this viewpoint. If reproductive isolation is considered as a decreased capacity for a type of nucleic acid transfer between organisms that can result in homologous recombination, then the classical reproductive isolation definition of species is applicable to all domains of life, including many that are often deemed asexual [24-26]. On this basis, if in some way homologous recombination between the nucleic acids of two individuals, be they cellular or acellular (i.e. viruses), is severely impaired, then they are likely to be members of different species. Indeed, recent ANI analyses indicate that while "the biological mechanisms underlying this genetic discontinuity are not clear," they "could involve a dramatic drop in recombination frequency around [below] 90–95% ANI, which could account for the discontinuity if bacteria evolve sexually … ." [27].  The sexual option is noted by Cohen [28] who recalls that with bacteria "laboratory studies have shown an exponential decay of genetic exchange rate with increasing sequence divergence, owing to DNA-sequence mismatch between organisms," however "recombination could maintain a high level of sequence identity among relatives but only for those with ANI >95%."

While it may be true for distantly related viruses that "viruses most likely do not have a single evolutionary origin and consequently lack any universal genes from which a shared genetics-based [i.e. gene-based] phylogeny could be constructed" [22], for closely related viruses this is less evident. Examples include various virus pairs that are likely to have a common ancestor and occupy a common host cell where recombination would occur should they not have evolved anti-recombination mechanisms (e.g. large k-mer differences) to maintain their reproductive isolation [29]. Thus, a finding that two co-infecting viruses have very different sequences [30] may not mean that one of them has evolved on a preferred host and is now coinfecting a less-preferred one [31].

Allied species of "viruses that infect the same species and cell types are thought to have evolved mechanisms to limit recombination" [32]. Thus, when we compare two viral species that have a *common* host cell, with two viral species that, even within a common *host*, do not share a common *cell*, we would expect to observe a fundamental difference related to their reproductive isolation mechanisms. If that difference is found to apply to other viral pairs that occupy a common host cell, then a fundamental isolation mechanism may have been identified [29]. Retroviruses are a good example [18, 33].

The extreme divergence of HIV1 and HTLV1 would initially have been favoured by small differences in ancestral GC% that would begin to impair recombination so reducing the genome blending which otherwise would have prevented establishment of independent populations. This mismatch form of reproductive isolation (analogous to chromosome-based hybrid sterility in sexual eukaryotes) had not at that stage been superseded by a gene-difference-driven form of reproductive isolation (either prezygotic malfunction, or post-zygotic hybrid inviability), as occurs in many sexual organisms. Thus, the nascent retroviral species were driven to GC% extremes, one to the AT-extreme (HIV1) and the other to the GC-extreme (HTLV1). Their descendants today are 'living fossils' to the extent that this fundamental form of reproductive isolation has been retained. In these viruses there has been no opportunity for gene-based reproductive isolation mechanisms to supersede chromosomal mechanisms [34, 35]. Newly arising incompatibilities between diffusible gene products have been unable to better this preexisting isolation mechanism, and there is no equivalent of prezygotic isolation as conventionally understood [2, 18].

Beyond all this, the 'metagenomics' revolution has forced taxonomists to rethink basic assumptions. Application of alignment-free technology to total DNA extracted from complex environments (e.g. sea water) has resulted in the description of many new species for which potential phenotypic characters can only be assessed indirectly from genotypic information. Indeed, noting that the description of "viruses known only by their sequence data continues to expand almost exponentially," Simmonds and Aiewsakun [36] suggest that taxonomists now pause "to discuss future species definitions for viruses, to critically evaluate the various biological species definitions currently in use, and to decide which concepts are most suitable for viruses in the future."

## Ad hoc DNA-based classification methods

Members of a species that are, by definition, not reproductively isolated from each other, are able to recombine. Successful recombination begins with the hybridization of aligned complementary nucleic acid sequences. This *in vivo* biological hybridization, reflecting closeness as a species, seems to have been simulated by the early DNA-DNA reassociation (DDR) test that depended on the alignment of DNAs purified from different (initially bacterial) sources; co-species membership was deemed to require >70% hybridization [37]. In contrast, Meier-Kolthoff *et al*. [38] find that "when inferred from genome sequences, within-species differences in the G+C content are almost exclusively below 1%" (i.e. co-species members are >99% similar).

In recent times the cumbersome DDR test has been replaced by various alignment-free procedures that have evolved from Chargaff's 'GC% rule' into the current ANI version with cut-off points much higher than 70% [1]. Simply stated, of 16 (= $4^2$) possible 2-mers, a high GC% species will have a high frequency of G-rich and C-rich 2-mers (GG, GC, CC, CG). A low GC% species will have a high frequency of A-rich and T-rich 2-mers (AA, AT, TT, TA). Intermediate GC% species will have more of the remaining eight 2-mers. The same reasoning applies to k-mers of higher orders (i.e. the 64 3-mers, the 256 4-mers, the 1024 5-mers, etc.).

The choice of appropriate k-mers was initially pragmatic. In 1993 Sitnikova and Sharkikh noted [39]:

"L-plet [k-mer] frequencies are significantly different for various species, and while mononucleotides allow one to distinguish only the main eubacterial groups, trinucleotides are genome-specific in both organelles and eubacteria. This makes it possible to use L-plet [k-mer] frequencies for estimation of evolutionary relatedness of species."

In 1995 Forsdyke [12] – following the linear regression approaches of Rogerson [40] and Prabhu [41] – used trinucleotides (3-mers) to distinguished species based on regression coefficient or slope values. With successively higher k-mer levels, Sandberg *et al*. [42] found major improvement up to the 3-mer level, then much slower improvement. For prokaryotes Richter and Rosselló-Móra in 2009 considered tetranucleotides (4-mers) a "genomic gold standard" [15]. This also works well for viruses [43].

Although a human first noticed the 1-mer connection to species [10], the hand of evolution might have first 'noticed' a higher k-mer connection, from which the 1-mer relationship would have then derived [12]. Indeed, an answer to the question as to whether one range of k-mers could be predicted from others, might indicate at what k-mer level biological selection pressures *had first acted*. This might help determine the nature of the underlying pressure. The question was first addressed from a mainly mathematical perspective. Along the lines of previous Markov chain analyses [44], which had shown that *higher* order k-mers could be predicted from *shorter* k-mers (order 3, 4), it was shown that the latter would predict even shorter k-mers (e.g. base composition [45]. In other words, if base composition (1-mer frequency) does not determine the frequency of trinucleotides and tetranucleotides (3-mers and 4-mers), and if higher order k-mers can be predicted from these, then 3-mers and 4-mers are likely to be *primary* in an evolutionary sense. This would be closely consistent with an anticodon-codon recognition analogy, where the 3-mer codon is supported by flanking bases such that there is near 5-mer recognition [46], which is in keeping with John Shepherd's RNY reading frame 'rule' [2, 47].

Those engaged in alignment-free analyses rightly take pleasure in pointing to the abundance of multiple full-length sequences that are now available. However, it has long been known that k-mer frequency is a genome-wide character, so that *genome fragments alone* suffice for many purposes. Thus, the complete genomes of members of individual species can be assembled from the sequences of mixtures of genome fragments obtained from the *total* DNA extracted from biological communities in a given environment ('metagenomics'). Indeed, from the analyses of the *incomplete* genomic sequences that became available in the 1970s, the 'father' of modern bioinformatics, Richard Grantham, was able to draw conclusions that stand today. Low coverage "genome skimming" can serve many taxonomic purposes [48]. So, while the number of complete sequences increasing exponentially in data-bases is welcome, the pressure for better bioinformatic methods to deal with this data should not divert attention from early interpretations, such as those of Grantham (see below), that may assist development of biologically relevant theory.

Furthermore, the alignment-free approach, which provides a metric for *entire* genomes, has cast doubt on deducing the timing and order of speciation events from phylogenetic trees constructed from *local* genetic data (i.e. from the sequences of individual or multiple gene loci). Classical phenotypic characteristics of organisms that are encoded by these loci are increasingly seen as of less importance. For prokaryotes, Konstantinidis and Tiedje [19] note "The combined data … reveal that two-thirds of the strains with 94% ANI differ in at least 5%, and up to 35%, of their total genes, revealing an extensive genetic diversity [i.e. gene diversity] *within a species*" (my italics). A reason for this is that speciation is usually a *process* of slow divergence during which gene flow can continue until full reproductive isolation is achieved. Thus, as noted by Nater *et al*. [49] "in such cases, a phylogenetic tree inferred from any given genomic locus (a 'gene' tree) might not unequivocally reflect the true order of speciation events (the 'species' tree), and inconsistent topologies are often obtained across the genome." Indeed, for prokaryotic populations individual genes may diverge quite independently of global sequence changes, where recombinational blending may thwart divergence into distinct species [50]. Such observations challenge both traditional 'genocentric' theory and attempts to "classify organisms into species based on gene flow" [51], as will be discussed below.

## Theoretical link with speciation

A mentor of Charles Darwin, with a chapter on the "origin and development of languages and species compared," drew attention to similarities between the divergence of ancestral species into new species, and the divergence of ancestral languages into new languages [52]. In the latter case, the earliest detectable changes were in accent (dialect). As colourfully set out in 1913 by Shaw in *Pygmalion* [53], the dropped Hs of Eliza separated her from Freddy. Under the tutelage of Professor Higgins this linguistic barrier was removed and, with it, their reproductive isolation. They lived happily ever after [16]. Since initially the barrier affected only a few words, there would have been many redundancies in a one-to-one alignment of their speeches. A comparison of the frequencies of certain H-containing words (i.e. non-alignment) would have been more informative.

Normally linguistic barriers are geographic and, given human mobility, it was unlikely that Eliza's London-based cockney accent would ever have matured into a distinct language. It seems that cockney English and conventional English will forever blend. On the other hand, the reproductive isolation necessary to prevent the blending of diverging incipient species can also be achieved, without geographical separation (i.e. sympatrically; [54, 55]), by means of internal genomic changes that may not directly involve genes (see below).

A linkage between information and heredity was drawn in the 1870s by Ewald Hering in Prague and by Samuel Butler in London who considered a chicken as merely an egg's way of making another egg. There is "an abiding memory between successive generations" wrote Butler in a popular novel [56], which preceded his four less-popular books on evolution [57]. However, the influence of information science on heredity became clearer in the 1940s when DNA emerged as the carrier of the abiding memory. Chargaff documented differences between organisms in the 1-mer frequencies (base compositions) of their DNAs. The four 'rules' he adduced facilitated the discovery, not only of the double helical structure of DNA (Chargaff's first parity rule), but also of a relationship between 1-mer frequencies (expressed as the percentage of the bases G and C) and species (Chargaff's GC% rule). A genome-wide uniformity of GC% was confirmed by Sueoka in 1961 [58] who noted that its frequency "is rather uniform not only among DNA molecules of an organism, but also with respect to different parts of a given molecule" [2, 59].

Although he referred to reproductively isolated "strains" rather than "species," Sueoka [58] suggested that there might be a link between, not only GC% and species, but also GC% and speciation:

"DNA base composition is a reflection of phylogenetic relationship. Furthermore, it is evident that those strains which mate with one another (i.e. strains within the same 'variety') have similar base compositions. Thus, strains of variety 1 ..., which are freely intercrossed have similar mean GC content. … If one compares the distribution of DNA molecules of *Tetrahymena* strains of different mean GC contents, it is clear that the difference in mean values is due to a rather uniform difference of GC content in individual molecules. In other words, assuming that strains of Tetrahymena have a common phylogenetic origin, when the GC content of DNA of a particular strain changes, all the molecules undergo increases or decreases of GC pairs in similar amounts."

If there was a 1-mer GC% correlation with species, then there might be stronger correlations at higher k-mer levels. If so, could such correlations be informative about the initiation of the speciation process, or would they merely be markers of the fact that the initiation had begun? From studies of eukaryotic 2-mer differences, in 1976 Subak-Sharpe and coworkers [60] inferred the existence of species specific (base order-dependent) "general designs" throughout the DNA of organisms; these imposed "constraints that are independent of polypeptide specifying function [i.e. genic function]." These observations were supported by Nussinov in 1981 [61] who noted complementarity between certain 2-mers (Chargaff's second parity rule) and identified selective constraint that is likely "structural in origin" and is "not for a better functional protein, but rather is primarily *for the DNA's own advantage*" (my italics).

Evolutionary forces acting at the 3-mer level are likely operative on entire genomes, though early studies were mainly focused on the gene-located 3-mer codons. From such studies Grantham in 1980 [62] was led to his "genome hypothesis," namely that "all genes in a genome … tend to have the same coding strategy" (i.e. show preference for certain codons). Furthermore, in agreement with Schaap's earlier account [63], he concluded that "mRNA sequences contain other information than that necessary for coding proteins. This other 'genome-type' information is mainly in the degenerate bases of the sequence. Consequently, it is largely *independent* of the amino acids coded" (my italics). From this non-gene-centric perspective he later pondered a phylogenetically important role for the "other information," namely in speciation [64]:

"What is the fundamental explanation for interspecific variation in coding strategy? Are we faced with a situation of continuous variation within and between species, thus embracing a Darwinian perspective of gradual separation of populations to form new species, of species to form new genera, etc.?"

A 4-mer level study in 1991 by Rogerson [9], built on earlier Markov analyses of chromosome segments of eukaryotes [8] and prokaryotes [44]. Referring to a k-mer as a "short sequence," he noted (my italics):

"These seemingly universal short sequence constraints could be contributors to many other patterns in DNA, especially the biased usage of codons within coding regions … . Codon bias in particular could be caused by the imposition of one structural design (the codons) *on top of a second design* (the short sequence distribution). This would rationalize the genome hypothesis of Grantham … . If sequence constraints vary from species to species and are dependent on a repetitive interaction of chromosomal DNA with cytoplasmic factor(s) [perhaps recombination-related proteins], the very presence of constraints might be related to *the process of speciation*."

## Molecular link with speciation

Alignment-based methods, such as BLAST, first find matching k-mers which act to 'seed' more extensive pairing interactions that can then extend, with gaps allowing correction for small insertions or deletions, provided there is broad sequence similarity [65]. This bioinformatic procedure resembles an early biological model for recombination. The DNA "unpairing postulate" of Francis Crick in 1971 [66] predicted that recombinational pairing of homologous DNAs would be initiated by short k-mers in the loops of stem-loop structures, which would be extruded from classical duplex helices when strands unpair. Short complementary 'kissing' duplex loops, like the well-established tRNA anticodon-codon k-mer duplexes [46], would seed a more extensive pairing interaction that would then extend until limited by lack of homology [67]. While dismissing this as "now only of historical interest," a biographer noted the *importance* Crick had attached to the long legend of a key figure illustrating his postulate in the journal *Nature*. Indeed, Crick had requested (unsuccessfully) that the Editor depart from usual practice and display the legend in a larger font [68].

Early studies with RNA [69], were extended in 1993 by Kleckner and Wiener [70], who reinvigorated the kissing loop idea for meiotic recombination, and its applicability has since been shown in a variety of homologous DNA pairing systems [71-74]. The exquisite sensitivity of the loop pairing to differences in 1-mers (and hence by the above arguments to higher order k-mers), was demonstrated by principle component analysis (PCA) in DNA folding studies [2, 75].

A genome-wide importance of short k-mers would be expected if they were critical to recombination, an internal process affecting both non-genic and genic regions. In the latter regions their recombinational role might be challenged by the evolutionary pressure to improve protein function (i.e. classical natural selection). Given the high importance of proteins it was easy to predict that the amino acid composition (AAC) of proteins would dominate and, by use of alternative codons, k-mer frequencies would be forced to adapt, as PCA shows they do to some extent [76, 77]. However, using PCA to explore alternative explanations for k-mer frequencies, in 2015 Brbić *et al*. [78] reported that, in general, *91% of the AAC variance* could be explained by "DNA-level processes" that were attributed to "directional mutational pressure," for which no further explanation was given. Furthermore, "despite contributing more to the effective proteome" genes with high expression levels did *not* differ in AAC from those with low levels. Thus, "evolutionary shifts in overall AAC appear to occur *almost exclusively* through factors shaping the *global* oligonucleotide content of the genome" (my italics).

One such factor is the purine-loading of open reading frames (ORFs) – a manifestation of Chargaff's 'cluster rule'[2]. This is confined to exons in eukaryotes, and can also affect AAC in prokaryotes [59, 79]. Another is the genome-wide pressure for recombination, which requires a dispersed potential to extrude stem-loop structures. Indeed, this may have engendered the original splitting of eukaryotic ORFs into exons and introns [77, 80]. Intriguing, in conflict with the thesis of Brbić *et al*. [78], when exons are under positive selection pressure to change rapidly, as in predator-prey 'arms races,' AAC appears to dominate over the structural demands of DNA. In this extreme circumstance, exonic stem-loop potential appears to divert into neighboring introns [81], which usually display k-mer frequencies ("linguistic constraints") like those of intergenic regions [76].

In this light, "directional mutational pressure" can be explained in terms of error-correction processes that require gene conversion [77, 82]. Facilitation of this would be an important role for meiotic recombination. Following a chromosomal anti-recombination speciation model [16, 29], the thwarting of such recombination by sequence changes (i.e. k-mer changes) that exceeded a certain threshold, would have played a continuing role, over long evolutionary periods, in the initiation of divergence into new species. In addition to the generally recognized aspects of an environment that foster *genic* evolution, an organism must contend with related organisms in that environment, which would foster *genomic* evolution [12]:

"Avoidance of recombination with organisms which have deviated from its own sequence (incipient and closely related species) is essential if an organism is to use recombination with organisms which have not deviated (members of its own species) as a means of maintaining the integrity of its own DNA. Species differences in (G+C)% (and hence differences in oligonucleotide hierarchies), would have arisen to impair stem-loop interactions, and hence impair recombination between species."

These GC% differences would reflect both mutational biases towards increasing or decreasing GC% and negative selection through attempted, but failed, recombination with related species.

## Nucleic acids versus proteins

The "DNA-level processes" of Brbić *et al*. [78] can be seen as operating internally on what has been referred to as the "genome phenotype" [83]. These processes can be distinguished from external influences that affect the conventional gene-governed phenotype, so changing mainly the first and second base positions of codons (non-synonymous mutations that change amino acids). The latter positions are prime targets of natural selection resulting in adaptations that affect biological fitness. However, a broader view of natural selection encompasses reproductive interactions that favor one genome over another and, when they operate in ORFs, change mainly the *third* base positions of codons (the synonymous mutations of Grantham's "degenerate bases"). Both types of natural selection can be adaptive in that they can increase the potential to produce fertile offspring – something argued since Darwin's time [2, 18]. Indeed, Venditti *et al*. in 2010 [84], from studies of the branch lengths of phylogenetic trees, excluded conventional natural selection as a general initiator of species divergence. Reminiscent of Romanes in 1897 [20], they attributed initiation of speciation to "rare stochastic events that cause reproductive isolation." A decoupling from speciation of adaptations likely to involve first and second codon positions was also noted in 2015 by Hedges and his colleagues [85].

This potential primacy of nucleic acids has been inferred from studies of codon choice in bacteria by Hershberg and Petrov [86]. They note that the "identity of favored codons tracks the GC content of the genomes," and consider that selection "appears to be consistently acting in the same direction as the nucleotide substitution bias of genomes." While they do not speculate how the "tracking" might occur, it is likely that when GC% begins to bias stochastically from the species norm, it will affect some genes before others. Their codons will no longer match their cell's tRNA pool. Genes encoding mRNAs whose rate of translation is not in any way rate-limiting will not initially need to change. But some genes, especially those that must rapidly increase expression in an emergency (e.g. genes encoding heat-shock proteins), will force tRNA pool adaptation. There will then be a selective pressure for increased expression of certain isoacceptor tRNAs (which bind the same amino acid but have a different codon), so changing the tRNAs available to *all* genes. Over time, the other genes will likely 'come into line' by accepting mutations that create codons that better match the transformed tRNA spectrum. Thus, 'GC-pressure' will be assisted by 'translation pressure.'

An interesting example of an environmental factor that can affect both proteins and nucleic acids is temperature, most dramatically expressed in thermophilic bacteria and archaea. Proteins isolated from these organisms are often thermostable and are valued commercially for their long shelf life. Thermophiles also display much higher purine-loading of ORFs than mesophiles [87]. Indeed, an AAC characteristic of thermophiles [88], includes many amino acids with purine-rich codons [89]. In keeping with this, Brbic *et al*. note [78] (my italics):

"A substantial component of the thermal AAC signature is grounded in oligonucleotide content, … . We obtain similar results when we try to discriminate halophiles from nonhalophiles. … Therefore, this analysis does not exclude selection on AAC in different environments, but implies that its signal is subtle when compared against the backdrop of the AAC changes dependent on oligonucleotide composition. … The key question then is whether the observed AAC changes are *purely a secondary effect* of the directional mutation pressures and/or adaptation of the DNA (or RNA) through oligonucleotide frequency shifts, while not necessarily being adaptive at the protein level."

Indeed, Dehouck *et al*. reported in 2008 [90] that the relationship between a protein's in vitro thermostability and the optimum growth temperature of the organism containing it, is not as close as previously thought. And in 2012, Liu *et al*. concluded [91] from a study of the purine-rich coding sequences of an enzyme that "the codons relating to enzyme thermal property are selected by thermophilic force at [the] nucleotide level," not at the protein level.

As viewed here, mutational pressure should not necessarily be continuously unidirectional. With some species there might be detectable biases towards increasing GC%. In others the biases might tend to decrease GC%. Over evolutionary time directions might change within a lineage [92]. There is also the problem of "inconsistent topologies" that generate taxonomic confusion [49]. For some it remains "a failure to correct for mutation pressure-driven compositional bias, present mainly in synonymous substitutions of protein-coding genes" that is "potentially a general source of phylogenetic artifacts" [93]. However, synonymous sites are likely to be the most reflective of genome-wide k-mer patterns that might more truly relate to phylogeny.

## Concluding remarks

This paper has had two major purposes. First to let information scientists and taxonomists know that there exists a possible biological basis for their alignment-free analyses. Second to inform biologists that, to the mounting evidence supporting chromosomal speciation [18, 94-98], should now be added that from alignment-free analyses. Given the complexities that make the bridging of these disciplines difficult [99], discussion of several relevant areas, such as horizontal gene transfer [5], convergent evolution and ribosomal RNA [37], has been omitted. Furthermore, usage of discipline-based technical terms has been minimized and, rather than paraphrase, authors (many well-recognized authorities in their fields) have been directly quoted. I conclude with an elementary outline that will hopefully be intelligible to students of various disciplines and to historians.

Speciation can occur in a *single lineage* in the sense that at some point new forms would arise that, if they could be tested with an ancestral form (created from fossil DNA *à la mode Jurassic Park* [100]), would be unable to cross to produce fertile offspring. The reproductive isolation that begins the speciation process [20] can occur in time (temporal isolation) or in space (geographical isolation). For our present purposes *divergent lineages* reflect speciation in the *same* time and space (in sympatry), because of processes internal to living creatures. Branching phylogenetic trees model this diversification. Traditionally, organisms showing similar characters are placed close on trees. Organisms showing different characters are placed at greater distances on trees. Some characters turn out to be more useful for tree construction than others. When nucleic acid sequences became available in the 1970s, researchers began to use them. The closer two sequences, the closer were considered the corresponding organisms.

Various methods for alignments of long strings of DNA bases were introduced. This facilitated the counting of the number of base differences between two sequences. However, the approach did not consider the possibility that some aspects of sequences, rather than long strings of bases, might better relate to the underlying evolutionary process that caused the species to diverge in the first place. When spoken languages begin to diverge there is first a difference in accents. In this case, lining up long texts would necessitate the inclusion of much redundant information. Some measure of accent difference might more efficiently display a relationship between the languages because redundancies would decrease.

This principle could partly underlie the success of alignment-free approaches that were introduced by information scientists without imputing a biological basis. In contrast to *local* gene-based disparities that biologists correctly argue can sometimes initiate divergence into species, it has here been suggested that *genome-wide* disparities in k-mer frequencies are the basis of sympatric chromosomal speciation – a most fundamental form of species initiation. When in conflict with its genes, a genome can sometimes win out [101].

**Keypoints**

- The alignment-free (k-mer) approach to sequence analysis seemed to have no underlying implications for biology when conceived by information scientists in the 1980s.

- Its success adds another pillar to arguments that the initiation of species divergence can have had a chromosomal, genome-wide, basis.

- A source of controversy among evolutionary biologists, these arguments have grown in strength since their elaboration in the 1990s.

- Underlying the new understanding are Chargaff's four 'rules,' Grantham's 'genome hypothesis,' and the evolutionary principles first glimpsed by Samuel Butler, George Romanes and William Bateson.

- Taxonomic categorizations, phylogenetic analyses, and species definitions that include microorganisms, now appear to rest on securer foundations.

**Acknowledgement**

Queen's University hosts my evolution webpages. The *arXiv* preprint server hosts an early version of this review. Thanks are also due to the Hypothesis Project which provided an accessible haven for comments when the NCBI removed links to *PubMed Commons* from *PubMed* items relevant to this work (e.g. PMID:26254668, PMID:27663721 and PMID:28877930); https://hypothes.is/search?q=tag%3APubMedCommonsArchive+forsdyke